\documentclass[11pt]{article}
\usepackage{amsmath,amsthm,amscd,array,amssymb}

\begin{document}

\begin{center}
{\Large{\bf $G_2$--invariant 7D Euclidean super Yang--Mills theory as a
\\
\medskip\smallskip
higher--dimensional analogue
of the 3D super--BF theory}}
\\
\bigskip\medskip
{\large{\sc D. M\"ulsch}}$^{a}$
\footnote{Email: muelsch@informatik.uni-leipzig.de}
and
{\large{\sc B. Geyer}}$^b$
\footnote{Email: geyer@itp.uni-leipzig.de}
\\
\smallskip
$\!\!\!\!\!^a$ Wissenschaftszentrum Leipzig e.V., D--04103 Leipzig, Germany
\\
\smallskip
{\it $^b$ Universit\"at Leipzig, Naturwissenschaftlich-Theoretisches Zentrum
\\
$~$ and Institut f\"ur Theoretische Physik, D--04109 Leipzig, Germany}

\end{center}

\begin{abstract}
\noindent {\small{A formulation of $N_T = 1$, $D = 8$ Euclidean super 
Yang--Mills theory with generalized self--duality and reduced 
$Spin(7)$--invariance is given which avoids the peculiar extra constraints 
of Ref. \cite{7}. Its reduction to seven dimensions
leads to the $G_2$--invariant $N_T = 2$, $D = 7$ super Yang--Mills theory 
which may be regarded as a higher--dimensional analogue of the 
$N_T = 2$, $D = 3$ super--BF theory. When reducing further that 
$G_2$--invariant theory to three dimensions one gets the $N_T = 2$ 
super--BF theory coupled a spinorial hypermultiplet.}}
\end{abstract}


\section{Introduction}

Recent developments in string duality and compactifications of
M--theory renewed the interest in those field theories in
dimensions $D > 4$ whose low energy effective action turns out to
be (essentially) that of dimensionally reduced supersymmetric
gauge theories. Such theories may arise naturally on the world
volumes of Euclidean D--branes wrapping manifolds of special holonomy. 
In particular, compactifications of $D = 11$ supergravity on compact
Joyce seven-- and eight--folds with $G_2$ and $Spin(7)$ holonomy,
respectively, have attracted some attention \cite{1}.

Moreover, independent of that development, it has been shown \cite{2,3,4} 
how the notion of topological quantum field theories \cite{5} or, more 
specifically, cohomological gauge theories being related to supersymmetric 
gauge theories by twisting, can be extended to dimensions greater than four. 
Such higher--dimensional (untwisted) cohomological theories are obtained 
when Euclidean supersymmetric gauge theories are considered on manifolds 
with reduced holonomy group $H \subset SO(D)$. These theories, which have 
a rather intriguing structure, localize onto the moduli space of certain 
generalized, higher--dimensional self--duality equations \cite{6}.

Recently, Euclidean super Yang--Mills theory (SYM) with generalized 
self--duality was explicitly established both in eight and seven dimensions 
with the $SO(8)$ and $SO(7)$ rotation invariance being broken down 
to $Spin(7)$ and $G_2$, respectively \cite{7}. In four dimensions the main 
ingredient connecting self--duality with simple or extended supersymmetry is
the chirality of fermions \cite{8}. In Ref. \cite{7} it has been
verified that in the case of generalized self--duality in $D > 4$
one needs, in addition to usual chirality, certain constraints 
on the supersymmetry parameters. Moreover, these extra constraints for 
the $G_2$--invariant theory can not be obtained from the $Spin(7)$--invariant 
theory in eight dimensions via simple dimensional reduction.

In this Letter we will re--analyse that problem, thereby relaxing the reality 
condition on fermions. We explicitly verify that when hermiticity in 
Euclidean space is abandoned --- which bears no problem here since 
hermiticity is primarily needed to ensure unitarity in Minkowski space --- 
chirality of fermions is a consistent and sufficient constraint 
being compatible with generalized self--duality, $Spin(7)$--invariance 
and octonionic algebra. Then, in fact, the $G_2$--invariant theory can be 
obtained by ordinary dimensional reduction. 
Moreover, that theory has a nice interpretation: It may be regarded as
the seven--dimensional analogue of the $N_T = 2$, $D = 3$ super--BF
theory \cite{9}, just as the $Spin(7)$--invariant theory may be regarded 
as the eight--dimensional analogue of the $N_T = 1$, $D = 4$
Donaldson--Witten theory \cite{3}. Namely, replacing in the $G_2$--invariant 
theory the octonionic through the quaternionic structure constants and 
considering all the fields as three--dimensional ones one gets exactly the 
$N_T = 2$, $D = 3$ super--BF theory (without matter). On the other hand, 
compactifying the $G_2$--invariant theory down to three 
dimensions gives the $N_T = 2$, $D = 3$ super--BF theory with matter.

\section{$Spin(7)$--invariant, $N_T = 1$, $D = 8$ Euclidean SYM}

Now, let us formulate the $Spin(7)$--invariant $N_T = 1$, $D = 8$
SYM without requiring the reality condition for the fermions.
Thereby, we avoid the subtlety of Ref.~\cite{7} which is associated 
with the compatibility of
dimensional reduction (from eight to seven dimensions) and generalized
self--duality (in seven dimensions). 

First, we introduce the $SO(8)$--invariant action of the Euclidean
$N = 2$, $D = 8$ SYM by ordinary dimensional reduction of the
Minkowskian $N = 1$, $D = 10$ SYM \cite{11} and subsequent Wick
rotation into the Euclidean space. Its field content consists of
an anti--hermitean vector field $A_a$ ($a = 1, \ldots, 8$),
16--component chiral and anti--chiral Weyl spinors, $\lambda$ and
$\bar{\lambda}$, respectively, and scalar fields $\phi$ and $\bar{\phi}$, 
all of them taking their values in the Lie algebra $Lie(G)$ of some 
compact gauge group $G$. As a result, one obtains
\begin{align}
\label{2.1}
S^{(N = 2)} = \int_E d^8x\, {\rm tr} \Bigr\{&
\hbox{$\frac{1}{4}$} F^{ab} F_{ab} +
2 \bar{\lambda} \Gamma^a D_a \lambda - 2 D^a \bar{\phi} D_a \phi
\nonumber \\ & 
+ 2 \lambda^T C_8^{-1} [ \phi, \lambda ] -
2 \bar{\lambda} C_8 [ \bar{\phi}, \bar{\lambda}^T ] -
2 [ \bar{\phi}, \phi ]^2 \Bigr\},
\end{align}
where $F_{ab} = \partial_{[a} A_{b]} + [ A_a, A_b ]$ and $D_a =
\partial_a + [ A_a, ~\cdot~ ]$ and $C_8$ is the charge conjugation
matrix, $C_8^{-1} \Gamma_a C_8 = - \Gamma_a^T$. 
For the moment, we do not specify the
$SO(8)$ matrices $\Gamma_a$ explicitly, $\frac{1}{2} \{ \Gamma_a,
\Gamma_b \} = \delta_{ab} I_{16}$.

The action (\ref{2.1}) is invariant under the following supersymmetry
transformations
\begin{align}
\label{2.2}
&\delta A_a = \bar{\zeta} \Gamma_a \lambda -
\bar{\lambda} \Gamma_a \zeta,
\nonumber
\\
&\delta \phi = \bar{\zeta} C_8 \bar{\lambda}^T,
\nonumber
\\
&\delta \bar{\phi} = \lambda^T C_8^{-1} \zeta,
\nonumber
\\
&\delta \lambda = - \hbox{$\frac{1}{4}$} \Gamma^{ab} \zeta F_{ab} +
\Gamma^a C_8 \bar{\zeta}^T D_a \bar{\phi} - \zeta [ \phi, \bar{\phi} ],
\nonumber
\\
&\delta \bar{\lambda} = \hbox{$\frac{1}{4}$} \bar{\zeta} \Gamma^{ab} F_{ab} -
\zeta^T C_8^{-1} \Gamma^a D_a \phi - \bar{\zeta} [ \bar{\phi}, \phi ].
\end{align}
Here, $\zeta$ and $\bar{\zeta}$ are constant chiral and anti--chiral Weyl
spinors, respectively, with
$\Gamma_9\, \zeta = - \zeta$, where $\Gamma_9 := \Gamma_1 \cdots \Gamma_8$, 
and $\Gamma_{ab} = \frac{1}{2} [ \Gamma_a, \Gamma_b ]$ are the generators
of $SO(8)$ rotations.

In order to get from (\ref{2.1}) a cohomological action with an
underlying $N_T = 1$ equivariantly nilpotent shift symmetry we
break down the Euclidean rotation group $SO(8)$ to $Spin(7)$,
i.e., we replace the $SO(8)$ matrices by the standard embedding
$\Gamma_a = ( \Gamma_A, \Gamma_8 )$, $A = 1,2, \ldots, 7$, of
$Spin(7)$ into $SO(8)$. In this representation 
we have (see, e.g., \cite{6})
\begin{equation}
\label{2.3}
\Gamma_A = \begin{pmatrix} 0 & - i (\gamma_A)_{\alpha\beta} \\
i (\gamma_A)_{\alpha\beta} & 0 \end{pmatrix},
\qquad
\Gamma_8 = \begin{pmatrix} 0 & \delta_{\alpha\beta} \\
\delta_{\alpha\beta} & 0 \end{pmatrix},
\qquad
\Gamma_9 = \begin{pmatrix} \delta_{\alpha\beta} & 0 \\
0 & - \delta_{\alpha\beta} \end{pmatrix},
\end{equation}
and for the charge conjugation matrix we may choose $C_8 = \Gamma_9$.
The 7 imaginary antisymmetric $Spin(7)$ matrices
$(\gamma_A)_{\alpha\beta}$ ($\alpha = 1, \ldots, 8$) are defined
by
\begin{equation*}
(\gamma_A)_{B8} = i \delta_{AB},
\qquad
(\gamma_A)_{BC} = i \Psi_{ABC},
\qquad
1 \leq A,B,C \leq 7,
\end{equation*}
with $\Psi_{ABC}$ being the $G_2$--invariant, totally antisymmetric
structure constants which enter into the division algebra of the
imaginary octonions \cite{12,13}. Their non--vanishing components are
\begin{equation*}
\Psi_{123} = \Psi_{246} = \Psi_{435} = \Psi_{367} =
\Psi_{651} = \Psi_{572} = \Psi_{714} = 1.
\end{equation*}
Besides, we introduce the dual octonionic structure constants, 
$\Phi_{ABCD} = - \hbox{$\frac{1}{6}$} \epsilon_{ABCDEFG}\, \Psi^{EFG}$, 
which, together with
\begin{equation}
\label{2.4}
\Phi_{8ABC} = \Psi_{ABC},
\end{equation}
define the $8$--dimensional $Spin(7)$--invariant self--dual Cayley tensor
$\Phi_{abcd} = \frac{1}{24} \epsilon_{abcdefgh} \Phi^{efgh}$.\footnote{
By the help of $\Phi_{abcd}$ one can {\it define} $Spin(7)$ 
as that subgroup of $SO(8)$ whose action on the
$8$--dimensional Euclidean space preserves the Cayley four--form
$\Phi = \hbox{$\frac{1}{24}$} \Phi_{abcd}\, d x^a \wedge d x^b
\wedge d x^c \wedge d x^d$.} 
The non--vanishing components of this tensor are
\begin{align*}
&\Phi_{1238} = \Phi_{2468} = \Phi_{4358} = \Phi_{3678}
= \Phi_{6518} = \Phi_{5728} = \Phi_{7148} = - 1,
\nonumber
\\
&\Phi_{4567} = \Phi_{3751} = \Phi_{6172} = \Phi_{5214}
= \Phi_{7423} = \Phi_{1346} = \Phi_{2635} = - 1.
\end{align*}
Below, we also need the following basic relations \cite{14},
\begin{align}
\label{2.5}
&\Psi^{ABE} \Psi_{CDE} = \delta^{[A}_{~~C} \delta^{B]}_{~~D} +
\Phi^{AB}_{~~~CD},
\nonumber
\\
&\Psi^{ABF} \Phi_{CDEF} = \hbox{$\frac{1}{2}$}
\Psi^{[A}_{~~[CD} \delta^{B]}_{~~E]},
\nonumber
\\
&\Phi^{ABCG} \Phi_{DEFG} =
\delta^{[A}_{~~D} \delta^B_{\!~~E} \delta^{C]}_{~~F} +
\hbox{$\frac{1}{4}$} \Phi^{[AB}_{~~~~[DE} \delta^{C]}_{~~F]} -
\Psi^{ABC} \Psi_{DEF}.
\end{align}

With the representation (\ref{2.3}) of the $SO(8)$ matrices
$\Gamma_a = ( \Gamma_A, \Gamma_8 )$ we get the
generators $\Gamma_{ab} = ( \Gamma_{AB}, \Gamma_{A8} )$ of the $SO(8)$
rotations as follows,
\begin{equation*}
\Gamma_{AB} = \begin{pmatrix} (\gamma_{AB})_{\alpha\beta} & 0 \\
0 & (\gamma_{AB})_{\alpha\beta} \end{pmatrix},
\qquad
\Gamma_{A8} = \begin{pmatrix} - i (\gamma_A)_{\alpha\beta} & 0 \\
0 & i (\gamma_A)_{\alpha\beta} \end{pmatrix},
\end{equation*}
where $(\gamma_{AB})_{\alpha\beta}$ are the antisymmetric $Spin(7)$ generators,
\begin{equation*}
(\gamma_{AB})_{C8} = \Psi_{ABC},
\qquad
(\gamma_{AB})_{CD} = \delta_{AC} \delta_{BD} - \delta_{AD} \delta_{BC} -
\Phi_{ABCD}.
\end{equation*}

Now, let us construct the $Spin(7)$--invariant action. First, we write 
the Weyl spinors as
\begin{equation*}
\lambda = - \Gamma_9 \lambda = \begin{pmatrix} 0 \\
\lambda_\alpha \end{pmatrix},
\qquad
\bar{\lambda} = \bar{\lambda} \Gamma_9 = ( \bar{\lambda}_\alpha, 0 ),
\end{equation*}
and consider, in Euclidean space, $\lambda_\alpha$ and $\bar{\lambda}_\alpha$ 
as two independent $8$--component spinors. Hence, $\lambda$ and 
$\bar{\lambda}$ are no longer subjected to the reality condition. 
More precisely, just as in Minkowski space, we take the adjoint spinor 
as $\bar{\lambda} = \lambda^\dagger \Gamma_8 = ( \lambda_\alpha^\dagger, 0 )$ 
and, afterwards, we drop the reality condition between $\lambda$ and 
$\bar{\lambda}$. In addition, for clarity, we change notation as 
$\lambda_\alpha^\dagger = \bar{\lambda}_\alpha$.

Next, we introduce a $Spin(7)$--octet of vector fields $\psi_a$, which is
obtained from the $8$--spinor $\bar{\lambda}_\alpha$ by identifying the
spinor index $\alpha$ with the vector index $a$, i.e., $\psi_a =
\bar{\lambda}_\alpha$ ($a, \alpha = 1, \ldots, 8$), a $Spin(7)$--septet
of self--dual tensor fields, $\chi_{ab} =
\frac{1}{6} \Phi_{abcd} \chi^{cd}$, and a $Spin(7)$--singlet scalar
field $\eta$, which are obtained from the $8$--spinor
$\lambda_\alpha = ( \lambda_A, \lambda_8 )$ according to
$\chi_{A8} = \lambda_A$, $\chi_{AB} = \Psi_{ACB} \lambda_C$ and
$\eta = \lambda_8$, respectively.

Then, after substituting in the action (\ref{2.1}) for $\Gamma_a$
the representation (\ref{2.3}), replacing $C_8$ through
$\Gamma_9$, and introducing the fields $\eta$, $\psi_a$ and
$\chi_{ab}$, one gets the following $Spin(7)$--invariant action,
\begin{align}
\label{2.6}
S_{\bigr| Spin(7) \subset SO(8)}^{(N_T = 1)} =
\int_E d^8x\, {\rm tr} \Bigr\{&
\hbox{$\frac{1}{4}$} F^{ab} F_{ab} -
2 D^a \bar{\phi} D_a \phi - 2 \chi^{ab} D_a \psi_b + 2 \eta D^a \psi_a
\nonumber
\\
& + 2 \bar{\phi} \{ \psi^a, \psi_a \} +
\hbox{$\frac{1}{4}$} \phi \{ \chi^{ab}, \chi_{ab} \} +
2 \phi \{ \eta, \eta \} - 2 [ \bar{\phi}, \phi ]^2 \Bigr\}.
\end{align}

Furthermore, from (\ref{2.2}), decomposing the (anti)chiral spinors
$\zeta$ and $\bar{\zeta}$ in the same manner as $\lambda$ and
$\bar{\lambda}$ and performing the same replacements as before,
after a straightforward but lengthy calculation, one gets the
following {\it on--shell} supersymmetry transformations:
\begin{alignat}{2}
\label{2.7}
&Q A_a = \psi_a,
&\qquad
&Q \psi_a = D_a \phi,
\nonumber
\\
&Q \phi = 0,
&\qquad
&Q \bar{\phi} = \eta,
\nonumber
\\
&Q \eta = [ \bar{\phi}, \phi ],
&\qquad
&Q \chi_{ab} = \hbox{$\frac{1}{4}$} \Theta_{abcd} F^{cd},
\\
\nonumber
\\
\label{2.8}
&Q_a A_b = \delta_{ab} \eta + \chi_{ab},
&\qquad
&Q_a \psi_b = F_{ab} - \hbox{$\frac{1}{4}$} \Theta_{abcd} F^{cd} +
\delta_{ab} [ \phi, \bar{\phi} ],
\nonumber
\\
&Q_a \phi = \psi_a,
&\qquad
&Q_a \bar{\phi} = 0,
\nonumber
\\
&Q_a \eta = D_a \bar{\phi},
&\qquad
&Q_a \chi_{cd} = \Theta_{abcd} D^b \bar{\phi}
\\
\intertext{and}
\label{2.9}
&Q_{ab} A_c = - \Theta_{abcd} \psi^d,
&\qquad
&Q_{ab} \psi_c = \Theta_{abcd} D^d \phi,
\nonumber
\\
&Q_{ab} \phi = 0,
&\qquad
&Q_{ab} \bar{\phi} = \chi_{ab},
\nonumber
\\
&Q_{ab} \eta = - \hbox{$\frac{1}{4}$} \Theta_{abcd} F^{cd},
&\qquad
&Q_{ab} \chi_{cd} = \hbox{$\frac{1}{4}$} \Theta_{abeg}
\Theta_{cdf}^{~~~~\!g} F^{ef} + \Theta_{abcd} [ \bar{\phi}, \phi ].
\end{alignat}
Here, we have introduced the (unnormalized) projector
\begin{equation}
\label{2.10}
\Theta_{abcd} = \delta_{ac} \delta_{bd} - \delta_{ad} \delta_{bc} +
\Phi_{abcd},
\qquad
\hbox{$\frac{1}{8}$} \Theta_{abef} \Theta_{cd}^{~~~\!ef} = \Theta_{abcd},
\end{equation}
which projects any antisymmetric second rank tensor onto its self--dual part
$\mathbf{7}$, according to the decomposition 
$\mathbf{28} = \mathbf{7} \oplus \mathbf{21}$ of the adjoint representation 
of $SO(8) \sim SO(8)/Spin(7) \otimes Spin(7)$.

In order to verify that the transformations (\ref{2.7}) -- (\ref{2.9}) really
leave the action (\ref{2.6}) invariant, one needs the following two
identities,
\begin{align}
\label{2.11}
&\hbox{$\frac{1}{2}$} ( \Theta_{abeg} \Theta_{cdf}^{~~~~\!g} -
\Theta_{abfg} \Theta_{cde}^{~~~~\!g} ) =
- \Theta_{efac} \delta_{bd} + \Theta_{efad} \delta_{bc} +
\Theta_{efbc} \delta_{ad} - \Theta_{efbd} \delta_{ac}
\nonumber
\\
&\phantom{\hbox{$\frac{1}{2}$} ( \Theta_{abeg} \Theta_{cdfg} -
\Theta_{abfg} \Theta_{cdeg} ) =,}
+ \Theta_{abce} \delta_{df} - \Theta_{abde} \delta_{cf} -
\Theta_{abcf} \delta_{de} + \Theta_{abdf} \delta_{ce}
\nonumber
\\
&\phantom{\hbox{$\frac{1}{2}$} ( \Theta_{abeg} \Theta_{cdfg} -
\Theta_{abfg} \Theta_{cdeg} ) =,}
- \Theta_{cdae} \delta_{bf} + \Theta_{cdbe} \delta_{af} +
\Theta_{cdaf} \delta_{be} - \Theta_{cdbf} \delta_{ae},
\nonumber
\\
&\hbox{$\frac{1}{2}$} ( \Theta_{abeg} \Theta_{cdf}^{~~~~\!g} +
\Theta_{abfg} \Theta_{cde}^{~~~~\!g} ) =
\Theta_{abcd} \delta_{ef},
\end{align}
which encode all of the algebraic properties of the structure constants
$\Psi_{ABC}$ and $\Phi_{ABCD}$ being displayed in Eqs. (\ref{2.4}) and 
(\ref{2.5}). Moreover, by the help of (\ref{2.11}), one can check that the 
transformations (\ref{2.7}) -- (\ref{2.9}) satisfy the following superalgebra 
{\it on--shell},
\begin{align*}
&\{ Q, Q \} \doteq - 2 \delta_G(\phi),
\qquad
\{ Q, Q_a \} \doteq \partial_a + \delta_G(A_a),
\qquad
\{ Q_a, Q_b \} \doteq - 2 \delta_{ab} \delta_G(\bar{\phi}),
\nonumber
\\
&\{ Q, Q_{cd} \} \doteq 0,
\qquad
\{ Q_a, Q_{cd} \} \doteq \Theta_{abcd} ( \partial^b + \delta_G(A^b) ),
\qquad
\{ Q_{ab}, Q_{cd} \} \doteq - 2 \Theta_{abcd} \delta_G(\phi),
\end{align*}
where $\delta_G(\varphi)$ denotes a gauge transformation with
field--dependent parameter $\varphi = (A_a, \phi, \bar{\phi})$ being
defined by $\delta_G(\varphi) = - D_a \varphi$ and
$\delta_G(\varphi) X = [ \varphi, X ]$ for all the other fields.
(The symbol $\doteq$ means that the corresponding relation is fulfilled
only {\it on--shell}.)

Finally, by adding to (\ref{2.6}) a topological term,
\begin{equation*}
S^{(N_T = 1)} = S_{\bigr| Spin(7) \subset SO(8)}^{(N_T = 1)} +
\int_E d^8x\, {\rm tr} \Bigr\{
\hbox{$\frac{1}{8}$} \Phi^{abcd} F_{ab} F_{cd} \Bigr\},
\end{equation*}
for the cohomological action we are looking for, by virtue of (\ref{2.10}),
one immediately obtains
\begin{align}
\label{2.12}
S^{(N_T = 1)} = \int_E d^8x\, {\rm tr} \Bigr\{&
\hbox{$\frac{1}{8}$} \Theta^{abcd} F_{ab} F_{cd} -
2 D^a \bar{\phi} D_a \phi - 2 \chi^{ab} D_a \psi_b + 2 \eta D^a \psi_a
\nonumber
\\
& + 2 \bar{\phi} \{ \psi^a, \psi_a \} +
\hbox{$\frac{1}{4}$} \phi \{ \chi^{ab}, \chi_{ab} \} +
2 \phi \{ \eta, \eta \} - 2 [ \bar{\phi}, \phi ]^2 \Bigr\}.
\end{align}
Except for some field rescalings, this is precisely the action given in
Refs. \cite{4}. Notice that, by virtue of
$\frac{1}{8} \Theta_{abef} \Theta_{cd}^{~~~\!ef} = \Theta_{abcd}$, only
the self--dual part $F_{ab}^+ = \frac{1}{8} \Theta_{abcd} F^{cd}$ enters
into the first term of (\ref{2.12}).

On--shell, upon using the equation of motion of $\chi^{ab}$, the action
(\ref{2.12}) can be recast into the $Q$--exact form $Q \Psi$, with the
gauge fermion
\begin{equation*}
\Psi = \int_E d^8x\, {\rm tr} \Bigr\{
\chi^{ab} ( F_{ab} - \hbox{$\frac{1}{16}$} \Theta_{abcd} F^{cd} ) -
2 \psi^a D_a \bar{\phi} - 2 [ \eta, \bar{\phi} ] \phi \Bigr\}.
\end{equation*}
Here, the first term enforces the localization onto the moduli space and
the third term ensures that pure gauge degrees of freedom are projected
out. The second and fourth term are typically for the Feynman type
gauge; they could be omitted, leading to the Landau type gauge.

\section{$G_2$--invariant, $N_T = 2$, $D = 7$ SYM
with global $SU(2)$ symmetry}

After having established the $Spin(7)$--invariant $N_T = 1$, $D = 8$ SYM ---
without introducing extra constraints --- the $G_2$--invariant
$N_T = 2$, $D = 7$ SYM can be simply obtained by ordinary dimensional
reduction.

First, we introduce two $SU(2)$--doublets of scalar and vector
fields, $\eta^\alpha$ and $\psi_A^{\!~~\alpha}$ ($\alpha = 1,2$), and a
$SU(2)$--triplet of scalar fields $G^{\alpha\beta}$,
\begin{equation*}
\eta^\alpha = \begin{pmatrix} \psi_8 \\ \eta \end{pmatrix},
\qquad
\psi_A^{\!~~\alpha} = \begin{pmatrix} \psi_A \\
\chi_{A8} = - \frac{1}{6} \Psi_{ABC} \chi^{BC} \end{pmatrix},
\qquad
G^{\alpha\beta} = \begin{pmatrix} \phi & \frac{1}{2} A_8 \\
\frac{1}{2} A_8 & \bar{\phi} \end{pmatrix},
\end{equation*}
where the spinor fields $\eta^\alpha$ and $\psi_A^{\!~~\alpha}$
are singlets and septets of the group $G_2$,
respectively.\footnote{ 
By the help of $\Psi_{ABC}$ one can {\it
define}  $G_2$ as the subgroup of $SO(7)$ whose action on
the $7$--dimensional Euclidean space preserves the associative
3--form $\Psi = \hbox{$\frac{1}{6}$} \Psi_{ABC}\, d x^A \wedge d
x^B \wedge d x^C$. Alternatively, it can be also characterized as
the maximal common subgroup of $SO(7)$ and $Spin(7)$, i.e., $G_2 =
SO(7) \cap Spin(7)$ \cite{12}.}\break The internal group index
$\alpha$ is raised and lowered as follows: $\varphi^\alpha =
\epsilon^{\alpha\beta} \varphi_\beta$ and $\varphi_\alpha =
\varphi^\beta \epsilon_{\beta\alpha}$, with
$\epsilon^{\alpha\gamma} \epsilon_{\beta\gamma} =
\delta^\alpha_{\beta}$, where $\epsilon^{\alpha\beta}$ is the
antisymmetric invariant tensor of the group $SU(2)$.

Then, by dimensional reduction, from (\ref{2.12}) one arrives at the
following $G_2$--invariant cohomological action with an underlying
$N_T = 2$ equivariantly nilpotent shift symmetry $Q^\alpha$ and global
symmetry group $SU(2)$,
\begin{align}
\label{3.1}
S^{(N_T = 2)} = \int_E d^7x\, {\rm tr} \Bigr\{&
\hbox{$\frac{1}{8}$} \Theta^{ABCD} F_{AB} F_{CD} -
D^A G_{\alpha\beta} D_A G^{\alpha\beta} -
\Psi^{ABC} \psi_{A \alpha} D_B \psi_C^{\!~~\alpha} -
2 \eta_\alpha D^A \psi_A^{\!~~\alpha}
\nonumber
\\
& + 2 G_{\alpha\beta} \{ \psi^{A \alpha}, \psi_A^{\!~~\beta} \} +
2 G_{\alpha\beta} \{ \eta^\alpha, \eta^\beta \} +
[ G_{\alpha\beta}, G_{\gamma\delta} ]
[ G^{\alpha\beta}, G^{\gamma\delta} ] \Bigr\}.
\end{align}
Here, analogous to (\ref{2.10}), we have introduced the (unnormalized)
projector (c.f., Eq.~(\ref{2.5}))
\begin{equation*}
\Theta_{ABCD} = \Psi_{ABE} \Psi_{CD}^{~~~~E},
\qquad
\hbox{$\frac{1}{6}$} \Theta_{ABEF} \Theta_{CD}^{~~~~EF} = \Theta_{ABCD},
\end{equation*}
which projects any antisymmetric second rank tensor onto its self--dual part
$\mathbf{7}$, according to the decomposition
$\mathbf{21} = \mathbf{7} \oplus \mathbf{14}$ of
$Spin(7) \sim Spin(7)/G_2 \otimes G_2$.

Next, we put $Q$, $Q_a = (Q_A, Q_8)$ and
$Q_{ab} = (Q_{AB}, Q_{A8} = - \frac{1}{6} \Psi_{ABC} Q^{BC})$ into the
following $SU(2)$--doublets of scalar and vector supercharges,
\begin{equation*}
Q^\alpha = \begin{pmatrix} Q \\ - Q_8 \end{pmatrix},
\qquad
Q_A^{\!~~\alpha} = \begin{pmatrix} - Q_{A8} \\ Q_A \end{pmatrix}.
\end{equation*}
Once more, performing the same dimensional reduction, from
(\ref{2.7}) -- (\ref{2.9}) one obtains the following {\it on--shell}
supersymmetry transformations:
\begin{align}
\label{3.2}
&Q^\alpha A_A = \psi_A^{\!~~\alpha},
\nonumber
\\
&Q^\alpha \psi_A^{\!~~\beta} = D_A G^{\alpha\beta} -
\hbox{$\frac{1}{4}$} \epsilon^{\alpha\beta} \Psi_{ABC} F^{BC},
\nonumber
\\
&Q^\alpha \eta^\beta = - \epsilon_{\gamma\delta}
[ G^{\alpha\gamma}, G^{\beta\delta} ],
\nonumber
\\
&Q^\alpha G^{\beta\gamma} = \hbox{$\frac{1}{2}$}
\epsilon^{\alpha(\beta} \eta^{\gamma)},
\\
\intertext{and}
\label{3.3}
&Q_A^{\!~~\alpha} A_B = \delta_{AB} \eta^\alpha -
\Psi_{ABC} \psi^{C \alpha},
\nonumber
\\
&Q_A^{\!~~\alpha} \psi_B^{\!~~\beta} = - \epsilon^{\alpha\beta} F_{AB} +
\hbox{$\frac{1}{4}$} \epsilon^{\alpha\beta} \Theta_{ABCD} F^{CD} +
\Psi_{ABC} D^C G^{\alpha\beta} +
\delta_{AB} \epsilon_{\gamma\delta} [ G^{\alpha\gamma}, G^{\beta\delta} ],
\nonumber
\\
&Q_A^{\!~~\alpha} \eta^\beta = D_A G^{\alpha\beta} -
\hbox{$\frac{1}{4}$} \epsilon^{\alpha\beta} \Psi_{ABC} F^{BC},
\nonumber
\\
&Q_A^{\!~~\alpha} G^{\beta\gamma} = - \hbox{$\frac{1}{2}$}
\epsilon^{\alpha(\beta} \psi_A^{\!~~\gamma)}.
\end{align}
By making use of (\ref{2.5}) one easily verifies that the above
transformations satisfy the following superalgebra {\it on--shell},
\begin{align*}
&\{ Q^\alpha, Q^\beta \} \doteq
- 2 \delta_G(G^{\alpha\beta}),
\qquad 
\{ Q^\alpha, Q_A^{\!~~\beta} \} \doteq
\epsilon^{\alpha\beta} ( \partial_A + \delta_G(A_A) ),
\\
&\{ Q_A^{\!~~\alpha}, Q_B^{\!~~\beta} \} \doteq
- 2 \delta_{AB} \delta_G(G^{\alpha\beta}) +
\epsilon^{\alpha\beta} \Psi_{ABC} ( \partial^C + \delta_G(A^C) ).
\end{align*}

Furthermore, on--shell, upon using the equation of motion for
$\psi_A^{\!~~\alpha}$, the action (\ref{3.1}) can be rewritten as
$S^{(N_T = 2)} \doteq \hbox{$\frac{1}{2}$}
\epsilon_{\alpha\beta} Q^\alpha Q^\beta \Omega$ with the gauge boson
\begin{equation}
\label{3.4}
\Omega = S_{\rm CS} + \int_E d^7x\, {\rm tr} \Bigr\{
\psi^A_{\!~~\alpha} \psi_A^{\!~~\alpha} - \eta_\alpha \eta^\alpha \Bigr\},
\end{equation}
where $S_{\rm CS}$ is the $7$--dimensional Chern--Simons action,
\begin{equation}
\label{3.5}
S_{\rm CS} = - \int_E d^7x\, {\rm tr} \Bigr\{
\Psi^{ABC} ( A_A \partial_B A_C + \hbox{$\frac{2}{3}$} A_A A_B A_C ) \Bigr\}.
\end{equation}

Let us notice that, with regard to the particular structure of
$\Omega$, in Ref. \cite{4} the question has been raised whether a
$7$--dimensional analogue of the Schwarz--type topological
Chern--Simons theory \cite{15} exists which can be obtained directly
from the action (\ref{3.5}). However, it was pointed out that, 
in contrast to the $3$--dimensional case, the quantization of such an action 
remains an open question because the Gauss law in the $A_7 = 0$ gauge is 
not sufficient to consistently solve the theory. Here instead, according to 
our construction, we observe that the action (\ref{3.5}) appears quite 
natural as the relevant part of the gauge fermion (\ref{3.4}) in the 
$G_2$--invariant Witten--type cohomological Yang--Mills theory.

\section{Dimensional reduction to three dimensions }

Independently, one may ask whether the $G_2$--invariant action
(\ref{3.1}) may be regarded as higher--dimensional analogue of
some topologically twisted action (in $D\leq 4$). In this section 
we show that the $N_T = 2$, $D = 3$ super--BF theory \cite{9}, 
with a spinorial hypermultiplet coupled to it, gets unified in the 
$G_2$--invariant action of the $N_T = 2$, $D = 7$ SYM.

In fact, identifying the octonionic structure
constants $\Psi_{ABC}$ for $1 \leq A,B,C \leq 3$ with the totally
antisymmetric Levi--Civita tensor $\epsilon_{ijk}$ ($i = 1,2,3$)
and considering all the fields as $3$--dimensional ones, we
recover from that part of (\ref{3.1}) precisely the {\it on--shell} 
formulation of the $N_T = 2$, $D = 3$ super--BF 
model with global symmetry group $SU(2)$,\footnote{
After introducing in (\ref{4.1}) an auxiliary field $B_i$ via 
its equation of motion,
$B_i = \frac{1}{4} \epsilon_{ijk} F^{jk}$, one recognizes the
usual {\it off--shell} formulation of the $N_T = 2$ super--BF
theory (without matter).}
\begin{align}
\label{4.1}
S_{\rm BF}^{(N_T = 2)} = \int_E d^3x\, {\rm tr} \Bigr\{&
\hbox{$\frac{1}{4}$} F^{ij} F_{ij} -
D^i G_{\alpha\beta} D_i G^{\alpha\beta} -
\epsilon^{ijk} \psi_{i \alpha} D_j \psi_k^{\!~~\alpha} -
2 \eta_\alpha D^i \psi_i^{\!~~\alpha}
\nonumber
\\
& + 2 G_{\alpha\beta} \{ \psi^{i \alpha}, \psi_i^{\!~~\beta} \} +
2 G_{\alpha\beta} \{ \eta^\alpha, \eta^\beta \} +
[ G_{\alpha\beta}, G_{\gamma\delta} ]
[ G^{\alpha\beta}, G^{\gamma\delta} ] \Bigr\}.
\end{align}

In addition, let us (formally) replace the structure constants
$\Psi_{iAB}$ and $\delta_{AB}$ for $4 \leq A,B \leq 7$ by $i
(\sigma_i)_{ab} \epsilon^{\alpha\beta}$ and $- \epsilon_{ab}
\epsilon^{\alpha\beta}$, respectively, $\sigma_i =
(\sigma_i)_a^{\!~~b}$ being the Pauli matrices, and let us put
$A_A$ and $\psi_A^{\!~~\alpha}$ for $4 \leq A \leq 7$ into the
$\overline{SU(2)}$--doublets $M_a^{~\alpha}$ and
$\lambda_a^{\!~~\alpha\beta}$ ($a = 1,2$), respectively. Then,
after dimensional reduction to three dimensions, from that 
remaining part of (\ref{3.1})
one gets the following action with global symmetry group $SU(2)
\otimes \overline{SU(2)}$,\footnote{
In order to recast the
(complex) matrices $i (\sigma_i)_{ab} \epsilon^{\alpha\beta}$ into the
(real) octonionic structure constants $\Psi_{iAB}$ one has to
perform an appropriate redefinition of $M_a^{~\alpha}$ and
$\lambda_a^{\!~~\alpha\beta}$.
}
\begin{align}
\label{4.2}
S^{(N_T = 2)} = S_{\rm BF}^{(N_T = 2)} + \int_E d^3x\, {\rm tr} \Bigr\{&
i \lambda_{a \alpha\beta} (\sigma^i)^{ab} D_i \lambda_b^{\!~~\alpha\beta} -
\hbox{$\frac{1}{2}$} D^i M_{a \alpha} D_i M^{a \alpha}
\nonumber
\\
& + 2 i \lambda_{a \alpha\beta} (\sigma^i)^{ab}
[ \psi_i^{\!~~\alpha}, M_b^{~\beta} ] +
2 \lambda_{a \alpha\beta} [ \eta^\alpha, M^{a \beta} ]
\nonumber
\\
& - 2 \epsilon_{\gamma\delta} G_{\alpha\beta}
\{ \lambda_a^{\!~~\alpha\gamma}, \lambda^{a \beta\delta} \} +
[ G_{\alpha\beta}, M_{a \gamma} ] [ G^{\alpha\beta}, M^{a \gamma} ] \Bigr\},
\phantom{\frac{1}{2}}
\end{align}
where the relations $(\sigma_i)_a^{\!~~c} (\sigma_j)_c^{\!~~b} =
\delta_{ij} \delta_a^{~b} + i \epsilon_{ijk} (\sigma^k)_a^{\!~~b}$ and
$(\sigma^i)_a^{\!~~b} (\sigma_i)_c^{\!~~d} =
\epsilon_{ac} \epsilon^{bd} - \delta_a^{~d} \delta_c^{~b}$ are used.
The action (\ref{4.2}) describes the $N_T = 2$, $D = 3$ super--BF model
with matter $(M_a^{~\alpha}, \lambda_a^{\!~~\alpha\beta})$ in the adjoint
representation. It may be regarded as the dimensionally reduced
non--Abelian version of the Seiberg--Witten (monopole) theory.

The full set of {\it on--shell} supersymmetry transformations which leave
(\ref{4.2}) invariant reads,
\begin{alignat*}{2}
&Q^\alpha A_i = \psi_i^{\!~~\alpha},
& \qquad 
&Q^\alpha \psi_i^{\!~~\beta} = D_i G^{\alpha\beta} -
\epsilon^{\alpha\beta} B_i,
\\
&Q^\alpha \eta^\beta = - \epsilon_{\gamma\delta}
[ G^{\alpha\gamma}, G^{\beta\delta} ],
&\qquad 
&Q^\alpha G^{\beta\gamma} = \hbox{$\frac{1}{2}$}
\epsilon^{\alpha(\beta} \eta^{\gamma)},
\\
&Q^\alpha M_a^{~\beta} = \lambda_a^{\!~~\alpha\beta},
&\qquad 
&Q^\alpha \lambda_a^{\!~~\beta\gamma} = - [ G^{\alpha\beta}, M_a^{~\gamma} ] -
\epsilon^{\alpha\beta} C_a^{~\gamma},
\\
\nonumber \\ 
&Q_i^{\!~~\alpha} A_j = \delta_{ij} \eta^\alpha -
\epsilon_{ijk} \psi^{k \alpha},
&\qquad 
&Q_i^{\!~~\alpha} \psi_j^{\!~~\beta} = - \epsilon^{\alpha\beta} F_{ij} +
\delta_{ij} \epsilon_{\gamma\delta} [ G^{\alpha\gamma}, G^{\beta\delta} ] +
\epsilon_{ijk} ( D^k G^{\alpha\beta} + \epsilon^{\alpha\beta} B^k ),
\\
&Q_i^{\!~~\alpha} \eta^\beta = D_i G^{\alpha\beta} -
\epsilon^{\alpha\beta} B_i,
&\qquad 
&Q_i^{\!~~\alpha} G^{\beta\gamma} = - \hbox{$\frac{1}{2}$}
\epsilon^{\alpha(\beta} \psi_i^{\!~~\gamma)},
\\
&Q_i^{\!~~\alpha} M_a^{~\beta} = - i (\sigma_i)_{ab} \lambda^{b \alpha\beta},
&\qquad 
&Q_i^{\!~~\alpha} \lambda_a^{\!~~\beta\gamma} = - \epsilon^{\alpha\beta}
D_i M_a^{~\gamma} - i (\sigma_i)_{ab} (
[ G^{\alpha\beta}, M^{b \gamma} ] - \epsilon^{\alpha\beta} C^{b \gamma} ),
\end{alignat*} 
{and}
\begin{align*}
&Q_a^{\!~~\alpha\beta} A_i = i (\sigma_i)_{ab} \lambda^{b \alpha\beta},
\\
&Q_a^{\!~~\alpha\beta} \psi_i^{\!~~\gamma} = \epsilon^{\alpha\gamma}
D_i M_a^{~\beta} + i (\sigma_i)_{ab} (
[ G^{\alpha\gamma}, M^{b \beta} ] - \epsilon^{\alpha\gamma} C^{b \beta} ),
\\
&Q_a^{\!~~\alpha\beta} \eta^\gamma = - [ G^{\alpha\gamma}, M_a^{\!~~\beta} ] -
\epsilon^{\alpha\gamma} C_a^{~\beta},
\\
&Q_a^{\!~~\alpha\beta} G^{\gamma\delta} = - \hbox{$\frac{1}{2}$}
\epsilon^{\alpha(\gamma} \lambda_a^{\!~~\delta)\beta},
\\
&Q_a^{\!~~\alpha\beta} M_b^{~\gamma} = - \epsilon_{ab}
\epsilon^{\beta\gamma} \eta^\alpha -
i (\sigma^i)_{ab} \epsilon^{\beta\gamma} \psi_i^{\!~~\alpha},
\\
&Q_a^{\!~~\alpha\beta} \lambda_b^{\!~~\gamma\delta} =
- \epsilon^{\alpha\gamma} [ M_a^{~\beta}, M_b^{~\delta} ] -
\epsilon_{ab} [ G^{\alpha[\beta}, G^{\delta]\gamma} ] +
i (\sigma^i)_{ab} \epsilon^{\beta\delta} (
D_i G^{\alpha\gamma} + \epsilon^{\alpha\gamma} B_i ),\qquad\qquad\qquad\quad
\end{align*}
with the abbreviations $B_i = \frac{1}{4} ( \epsilon_{ijk} F^{jk} +
i \epsilon_{\alpha\beta} (\sigma_i)_{ab} [ M^{a \alpha}, M^{b \beta} ] )$
and $C_a^{~\alpha} = - \frac{1}{2} i (\sigma^i)_{ab} D_i M^{b \alpha}$.
These transformations obey the following superalgebra {\it on--shell},
\begin{align*}
&\{ Q^\alpha, Q^\beta \} \doteq
- 2 \delta_G(G^{\alpha\beta}),
\qquad\quad 
\{ Q^\alpha, Q_i^{\!~~\beta} \} \doteq
\epsilon^{\alpha\beta} ( \partial_i + \delta_G(A_i) ),
\\
&\{ Q_i^{\!~~\alpha}, Q_j^{\!~~\beta} \} \doteq
- 2 \delta_{ij} \delta_G(G^{\alpha\beta}) +
\epsilon^{\alpha\beta} \epsilon_{ijk} ( \partial^k + \delta_G(A^k) ),
\\
&\{ Q^\alpha, Q_a^{\!~~\beta\gamma} \} \doteq
\epsilon^{\alpha\beta} \delta_G(M_a^{~\gamma}),
\qquad 
\{ Q_i^{\!~~\alpha}, Q_a^{\!~~\beta\gamma} \} \doteq
i \epsilon^{\alpha\beta} (\sigma_i)_{ab} \delta_G(M^{b \gamma}),
\\
&\{ Q_a^{\!~~\alpha\gamma}, Q_b^{\!~~\beta\delta} \} \doteq
2 \epsilon_{ab} \epsilon^{\gamma\delta} \delta_G(G^{\alpha\beta}) +
i \epsilon^{\alpha\beta} \epsilon^{\gamma\delta}
(\sigma^i)_{ab} ( \partial_i + \delta_G(A_i) ).
\end{align*}
On--shell, upon using the equations of motion for $\psi_i^{\!~~\alpha}$ and
$\lambda_a^{\!~~\alpha\beta}$, the action (\ref{4.2}) can be recast into
the form $S^{(N_T = 2)} \doteq \hbox{$\frac{1}{2}$} \epsilon_{\alpha\beta}
Q^\alpha Q^\beta \Omega$ with the gauge boson
\begin{equation*}
\Omega = S_{\rm CS} + \int_E d^3x\, {\rm tr} \Bigr\{
i (\sigma^i)^{ab} M_{a \alpha} D_i M_b^{~\alpha} +
\lambda^a_{\!~~\alpha\beta} \lambda_a^{\!~~\alpha\beta} +
\psi^i_{\!~~\alpha} \psi_i^{\!~~\alpha} - \eta_\alpha \eta^\alpha \Bigr\},
\end{equation*}
where $S_{\rm CS}$ is the $3$--dimensional Chern--Simons action \cite{15},
\begin{equation*}
S_{\rm CS} = - \int_E d^3x\, {\rm tr} \Bigr\{
\epsilon^{ijk} ( A_i \partial_j A_k +
\hbox{$\frac{2}{3}$} A_i A_j A_k ) \Bigr\}.
\end{equation*}

Summarizing, we have shown that, in the Euclidean space, when relaxing the 
reality constraint on fermions, then generalized self--duality, simple 
supersymmetry, $Spin(7)$ invariance and octonionic algebra are compatible 
with each other and with chirality in eight dimensions, just as 
self--duality and supersymmetry are compatible with usual chirality in 
four dimensions. Additionally, we have observed that the fields of the gauge 
and spinorial hypermultiplet of the $N_T = 2$, $D = 3$ super BF--theory 
with matter gets unified in the fields of the gauge multiplet of the
$G_2$--invariant $N_T = 2$, $D = 7$ super Yang--Mills theory.


\end{document}